\input ./style/arxiv-general.cfg
\documentclass[aoas,MSNbibl,nameyear,dvips]{arximspdf}
\makeatletter
   \@ifpackageloaded{graphicx}{}{\usepackage{graphicx}}
\makeatother
\usepackage{multirow,dcolumn}

\doi{10.1214/15-AOAS820}
\volume{9}
\issue{2}
\pubyear{2015}
\firstpage{950}
\lastpage{968}
\docsubty{FLA}

\makeatletter

\newcommand{\eqref}[1]{(\ref{#1})}
\newcolumntype{d}[1]{D{.}{.}{#1}}
\makeatother

\begin{document}
\begin{frontmatter}

\title{Bayesian detection of embryonic gene expression onset in \emph{C. elegans}\thanksref{T1}}
\thankstext{T1}{Supported in part by a grant from the Research Grants Council of
the Hong Kong SAR (Project no. CUHK 400913).}
\runtitle{Bayesian Detection of expression onset}
\pdftitle{Bayesian detection of embryonic gene expression onset in C. elegans}

\begin{aug}
\author[A]{\fnms{Jie} \snm{Hu}\thanksref{M1}\ead[label=e1]{hujiechelsea@gmail.com}},
\author[B]{\fnms{Zhongying} \snm{Zhao}\thanksref{M2}\ead[label=e2]{zyzhao@hkbu.edu.hk}},
\author[C]{\fnms{Hari Krishna} \snm{Yalamanchili}\thanksref{M3}\ead[label=e3]{hari@hku.hk}},
\author[C]{\fnms{Junwen}~\snm{Wang}\thanksref{M3}\ead[label=e4]{junwen@hku.hk}},
\author[D]{\fnms{Kenny} \snm{Ye}\thanksref{M4}\ead[label=e5]{kenny.ye@einstein.yu.edu}}
\and
\author[A]{\fnms{Xiaodan} \snm{Fan}\corref{}\thanksref{M1}\ead[label=e6]{xfan@sta.cuhk.edu.hk}\ead[label=u1,url]{http://www.sta.cuhk.edu.hk/xfan/}}
\runauthor{J. Hu et al.}
\affiliation{Chinese University of Hong Kong\thanksmark{M1},
Hong Kong Baptist University\thanksmark{M2},  University of
Hong Kong\thanksmark{M3} and Albert Einstein College of
Medicine\thanksmark{M4}}
\address[A]{J. Hu\\
X. Fan\\
Department of Statistics\\
Chinese University of Hong Kong\\
Shatin\\
Hong Kong\\
\printead{e1}\\
\phantom{E-mail:\ }\printead*{e6}\\
\printead{u1}}

\address[B]{Z. Zhao\\
Department of Biology\\
Hong Kong Baptist University\hspace*{26pt}\\
Kowloon Tong\\
Hong Kong\\
\printead{e2}}

\address[C]{H.~K. Yalamanchili\\
J. Wang\\
Department of Biochemistry\\
University of Hong Kong\\
Pokfulam\\
Hong Kong\\
\printead{e3}\\
\phantom{E-mail:\ }\printead*{e4}}

\address[D]{K. Ye\\
Department of Epidemiology\\
\quad and Population Health\\
Albert Einstein College of Medicine\\
1300 Morris Park Avenue\\
Bronx, New York 10461\\
USA\\
\printead{e5}}
\end{aug}

%
\received{\smonth{9} \syear{2014}}
%
\revised{\smonth{1} \syear{2015}}

%
\begin{abstract}
To study how a zygote develops into an embryo with different
tissues, large-scale 4D confocal movies of \emph{C. elegans}
embryos have been produced recently by experimental biologists.
However, the lack of principled statistical methods for the highly
noisy data has hindered the comprehensive analysis of these data
sets. We introduced a probabilistic change point model on the cell
lineage tree to estimate the embryonic gene expression onset time.
A Bayesian approach is used to fit the 4D confocal movies data to
the model. Subsequent classification methods are used to decide a
model selection threshold and further refine the expression onset
time from the branch level to the specific cell time level.
Extensive simulations have shown the high accuracy of our method.
Its application on real data yields both previously known results
and new findings.
\end{abstract}

%
\end{frontmatter}

\section{Introduction}\label{introduction}
The process of how a single-cell zygote develops into an embryo
with different tissues is still a fundamental but open problem in
biology. Undoubtedly, gene expression dynamics plays a key role in
this procedure. Understanding when and where a gene starts
expression in the embryo, that is, the embryonic gene expression
onset, is a crucial step for solving this puzzle.

Modern high throughput experimental techniques, such as microarray
experiments and time-lapse confocal microscopy, can produce gene
expression data with high spatial and temporal resolution, which
is necessary for the study of embryogenesis. \emph{C. elegans} is
often used as the model organism for embryogenesis study due to
its transparency and invariant cell lineage from zygote to adult
[\citet{Sulston1983}]. \citet{Bao2006} and \citet{Murray2008}
introduced a system to automatically analyze the continuous
reporter gene expression in \emph{C. elegans} with cellular
resolution from zygote to embryo using the confocal laser microscope.
With this automatic system, \citet{Murray2012} analyzed the
expression patterns of 127 genes and provided a compendium of gene
expression dynamics. \citet{Long2009} and \citet{Liu2009} also
developed an analyzer to convert high-resolution confocal laser
microscope images into data tables and then analyzed cell fate
from gene expression profiles. Later \citet{Spencer2011} took
advantage of a spatial and temporal map of \emph{C. elegans} gene
expression to provide a basis for establishing the roles of
individual genes in cellular differentiation.

The aforementioned confocal microscopy on \emph{C. elegans}
embryogenesis is for tracing the expression of one specific target
gene on an individual embryo. Due to strain differences (such as the
insertion of green fluorescent protein DNA sequence into different
locations of the \emph{C. elegans} genome) and variability in experimental and
environmental factors, even data sets for measuring the same gene show
high quantitative variation, indicating considerable noise.
Furthermore, the expression change on
the embryonic cell lineage poses a change point problem on a
binary tree, which is a nonlinear problem rarely studied by
current literatures. The lack of principled statistical methods
makes the comprehensive understanding of these data sets too crude
to be convincing. For example, \citet{Murray2012} used an ad hoc
threshold to report the expression onset among all the data sets,
which ignored the variation among different runs of confocal
microscopy. Here, we apply a Bayesian method for automatic
detection of gene expression onset from the 4D confocal
microscopy data by introducing experiment-specific background and
signal distributions, which in turn can benefit downstream
analysis, such as gene network inference based on such high
spatial and temporal data [\citet{Yalamanchili2013}].

\begin{figure}[t]
\centering
\begin{tabular}{@{}cc@{}}

\includegraphics{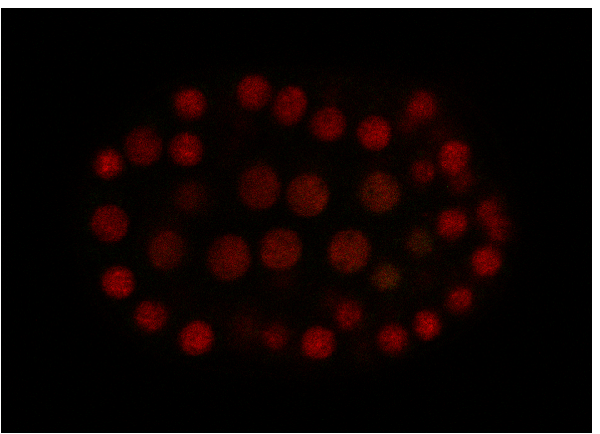}  & \includegraphics{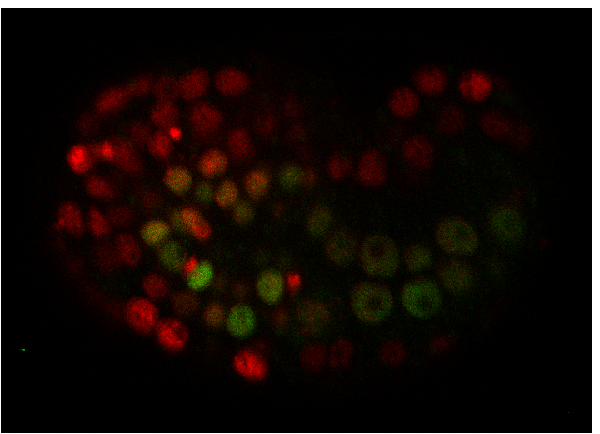}\\
\footnotesize{(A) 150-cell stage} & \footnotesize{(B) 550-cell stage}
\end{tabular}
\caption{Confocal fluorescent images of a \emph{C. elegans} embryo.
\textup{(A)} The embryo at the 150-cell stage, with ubiquitous labeling of
nuclei by red fluorescent protein mCherry; \textup{(B)} The embryo at the
550-cell stage, with ubiquitous labeling of nuclei by red
fluorescent protein mCherry and specific labeling of the gene
expression product of PHA-4 by green fluorescent protein. The
expression cells are in pharynx and intestine.} \label{f0}
\end{figure}

\begin{figure}[b]

\includegraphics{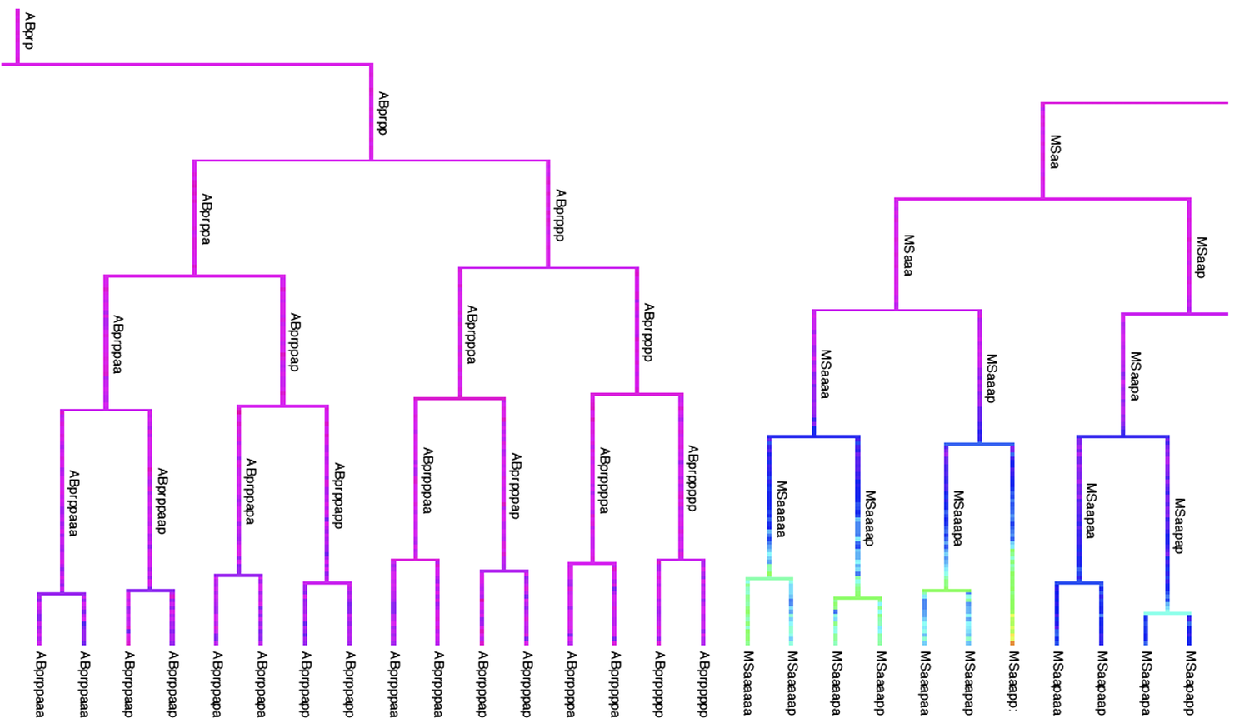}

\caption{An example of data. The figure shows a part of a cell lineage tree
for one data file, which represents the measured fluorescent
intensity from one time-lapse confocal microscopy experiment on
\emph{C. elegans}. The cell lineage and cell nomenclature are from
\citet{Sulston1983}.} \label{f1}
\end{figure}

Our real data application is based on the data provided by
\citet{Murray2012}, which is downloadable from
\url{http://epic.gs.washington.edu/}. Figure~\ref{f0} shows the confocal
fluorescent images of two stages of an embryo. The green fluorescent
protein labels the expression product of the gene PHA-4, which appears
in the 550-cell stage but not in the
150-cell stage. Figure~\ref{f1} shows a part of a cell lineage
tree from one data file, which corresponds to one run of the
confocal microscopy on one embryo. Each horizontal line represents
a cell division event. Each vertical line represents a cell with
the length proportional to its lifetime (i.e., how long a cell
lived). The color at each point of the vertical line represents
the measured fluorescent intensity at the corresponding time,
which gradually increases with the color changing from purple to
red as the rainbow color order. (In the later content, we use figures
with gray scales to represent tree structures and measured fluorescent
intensity gradually increases with the color changing from white to
black.) The blue and green cells in
Figure~\ref{f1} form a cluster whose fluorescence level is
significantly higher than the overall background, which indicates
that they may express the target gene. Thus, estimating
expression onset is actually a change point detection problem.
However, methods used to detect change points in regular
one-dimensional time series, such as
\citet{Guralnik1985,Guralnik1999} and \citet{Perreault2000}, cannot
handle the tree structure in our case.

In Section~\ref{methods} we present a four-step method to detect
the onset time, where the key step is a Bayesian algorithm to fit
a change point model to the tree data. We apply this method on
both synthesized data and real data and show the estimation
results in Section~\ref{results}. Section~\ref{conclusion}
concludes the paper. Other details of the algorithm and the model
diagnosis are provided in supplemental materials [\citet{Hu2015}].

\section{Methods}\label{methods}
We introduce the following four-step method for the Detection of
Embryonic Gene Expression Onset (DEGEO), where the key step features a
probabilistic change point model on the cell lineage tree and a full Bayesian
approach to infer the cell where a gene starts expressing:
\begin{itemize}
\item \textit{Step} 1: summarize the measured fluorescent intensity of
each cell to a single cell score.

\item \textit{Step} 2: fit the tree of cell scores to a
change-point-in-tree model in order to detect an expression
branch, where a Markov chain Monte Carlo (MCMC) algorithm is used
to estimate the change point and the other model parameters.

\item \textit{Step} 3: use Support Vector Regression (SVR) to decide when
to stop detecting extra expression branches.

\item \textit{Step} 4: refine the onset detection by detecting the specific
onset time on the reported expression branches.
\end{itemize}

\subsection{Experiment and data}
For each 4D confocal laser scanning microscope experiment
performed by \citet{Murray2012}, we have a data file containing a
time series for each embryonic cell from its birth to its division
or death. Each measurement is a fluorescent intensity at each time
point (on average, one data value per minute) over the duration of
the cell's life. We use the time series data of Column ``blot'' in
the data files downloaded from \url{http://epic.gs.washington.edu/},
which has been normalized in order to reduce the influence of
background noise. We represent this measured fluorescent intensity
data of the $i$th cell at the $j$th time point by $y_{ij}$.
Other details about the real data are provided in the \hyperref[app]{Appendix}.

\subsection{Assumptions}\label{ass}
During the embryogenesis process, once a cell initializes the
expression of a gene, its descendants will inherit some of this
gene's products and may also continue expressing this gene. Thus,
a positive correlation between relatives is expected. Therefore,
we make a transitivity assumption by assuming the following: if a cell
expresses a gene, its child cells will also express the corresponding
gene and the gene expression values of the two sibling cells are
positively correlated. This assumption is justified by the data as
shown in \hyperref[suppA]{Supplement A}.

Experimenters have two methods to mark an expressed gene, namely,
promoter fusion and protein fusion [\citet{Murray2008}]. A special
characteristic of the data from promoter fusion is that the
fluorescent protein degrades much more slowly than that of protein
fusion. Thus, once a cell
initializes a gene's expression, the resulting fluorescent
intensity will be inherited by its descendants and seldom decreases.
If the child cells continue expressing this gene, the fluorescent
intensity will increase due to the accumulation of the fluorescent
protein. Since most experiments are based on promoter fusion, we
assume a general nondecreasing trend for the fluorescent
intensities along the paths from ancestors to descendants on an
expression branch. Again, we use the data to justify this
assumption as shown in \hyperref[suppA]{Supplement A}. This paper focuses on data from
promoter fusion.

Furthermore, we assume that when a gene is not expressed in a
cell, its cell score, which is defined in
Section~\ref{SecStep1}, follows a normal distribution with
parameters $\mu$ and $\sigma_{1}^{2}$. And the histogram of two
control files are listed in \hyperref[suppA]{Supplement~A}.

\subsection{The DEGEO procedure}
\subsubsection{Step 1: Summarize the time series of a cell into one cell
score}\label{SecStep1} Due to the abnormal fluctuation of
$y_{ij}$ right before and after the cell division time, we
truncate the first two and last two data points for all cells
whose lifetimes are more than 8 time points (96.2\% of the cells belong
to this group). Cells with fewer data points are truncated less.
The remaining data points are called the valid data. We define the
cell score for each cell as
\[
x_{i}=\frac{y_{i}^{(0.05)}+y_{i}^{(0.95)}}{2},
\]
where $y_{i}^{(0.05)}$ and $y_{i}^{(0.95)}$ denote the $5\%$ and
$95\%$ quantiles of the time series $\{y_{ij}\}$ of the $i$th
cell, respectively. The cell score is designed in this way such that a
true expression signal (which should last longer than $5\%$ of the
cell's lifetime) could be captured even if the expression lasts shorter
than half of the cell's lifetime (in this case, taking median may not
discover the expression). On the other time, rare outlier values (which
should not occupy more than $5\%$ of the cell's lifetime) can be
filtered out from the cell score. In contrast, a median will miss short
trends while a mean will be too sensitive to outliers. Thus, a 4D
confocal movie data file is transformed to a tree of cell scores. The
cell scores $X$, together with the lifetimes $T$ and their family
relationships, will be used in step 2 to detect the cells where the
target gene start expressing.

\subsubsection{Step 2: Fit a change-point-in-tree model}\label{step2}
Let $x_{i_{1}}$ and $x_{i_{2}}$ be the cell scores of a pair of
sibling cells while $x_{i_{0}}$ indicates that of their mother
cell. Let $t_{i_{1}}$ and $t_{i_{2}}$ be the lifetimes of the
cells corresponding to $x_{i_{1}}$ and $x_{i_{2}}$. $M$ indicates
the change point, that is, the cell where the target gene starts
expressing. Therefore, all descendant cells of the cell $M$ form a
branch, which we call an expression branch. In the case that cells
with close kinship are expression onsets simultaneously, the
change point may be the most recent common ancestor cell of all
expression cells. In this case, the expression branch may contain
cells which have not expressed the target gene. For example, in
Figure~\ref{f10}, the exact expression onset cells are Exxx, but
our algorithm reports the cell E as change point in step 2.
Nevertheless, our algorithm will refine it to Exxx in step 4.

Denote $A(x_{i})$ as the set of all ancestor cells of the cell
corresponding to the cell score $x_{i}$. If a cell corresponding to
the cell score $x_{i}$ is not within the expression branch, that is,
$M\notin A(x_{i})$, we assume its cell score is independent and
identically distributed (i.i.d.) Gaussian noise. For a cell in the
expression branch, its cell score is assumed to be associated with
its mother, its sibling and its lifetime. More specifically,
the two kinds of cell scores are modeled by a
change-point-in-tree model as follows:
%
\begin{eqnarray}
\label{Model1}
&& x_{i}| M\notin A(x_{i}) \sim N\bigl(\mu,
\sigma_{1}^{2}\bigr),\nonumber
\\
&&\pmatrix{ x_{i_{1}}
\cr
x_{i_{2}}}\Big| x_{i_{0}}, M\in
A(x_{i_{1}}, x_{i_{2}}) \\
&&\qquad \sim N\left( \pmatrix{ x_{i_{0}}+\beta t_{i_{1}}
\vspace*{2pt}\cr
x_{i_{0}}+\beta t_{i_{2}}}, \pmatrix{ \sigma_{2}^{2}&\rho
\sigma_{2}^{2}
\vspace*{2pt}\cr
\rho\sigma_{2}^{2}&\sigma_{2}^{2}}\right).\nonumber
\end{eqnarray}

The above change-point-in-tree model contains one unknown change
point $M$ and five unknown parameters. We will use a Bayesian
approach to estimate them from a data file. To facilitate Bayesian
computing, we use conjugate prior distributions for unknown
parameters. Detail prior distributions are as follows:
\begin{eqnarray*}
\sigma_{1}^{2} &\sim&\Gamma^{-1}(g,h),
\\
\sigma_{2}^{2} &\sim&\Gamma^{-1}(a,b),
\\
\beta&\sim& N(r,s),
\\
\mu &\sim&N(p,q),
\\
\rho&\sim& \operatorname{Beta}(u,v),
\\
M &\sim&\mbox{Uniform over all cells in the candidate set}.
\end{eqnarray*}
Settings of the hyperparameters in the above prior distributions
and the sensitivity analysis are listed in \hyperref[suppB]{Supplement B}. The
candidate set contains all cells on the cell lineage tree which
may initialize a gene's expression. Here we exclude boundary cells
of the tree from the candidate set because an expression pattern
changing at the boundary is either false positive or a signal too
weak to be significantly different from the background. More
specifically, considering the situation of the \emph{C. elegans}
embryo, only cells whose number of descendants is between 6 and 30
are put in the candidate set, while for a large expression branch,
the DEGEO algorithm will divide it into several small expression
branches and detect them one by one. The joint posterior
distribution is as follows:
\begin{eqnarray*}
&&f\bigl(\sigma_{1}^{2},
\sigma_{2}^{2},\beta,\mu,\rho,M|X,T\bigr)\\
&&\qquad\propto f\bigl(
\sigma _{1}^{2},\sigma_{2}^{2},\beta,\mu,
\rho,M\bigr)\cdot f\bigl(X,T|\sigma _{1}^{2},
\sigma_{2}^{2},\beta,\mu,\rho,M\bigr)
\\
&&\qquad\propto\bigl(\sigma_{1}^{2}\bigr)^{-g-1}e^{-{h}/{\sigma_{1}^{2}}}
\cdot \bigl(\sigma_{2}^{2}\bigr)^{-a-1}e^{-{b}/{\sigma_{2}^{2}}}
\cdot e^{-
{(\beta-r)^{2}}/{(2s)}}\\
&&\qquad\quad{}\times e^{-{(\mu-p)^{2}}/{(2q)}}\cdot\rho ^{u-1}(1-
\rho)^{v-1}
\\
&&\qquad\quad{}\times\frac{1}{\sigma_{1}^{|N_{M}|}}e^{-({1}/{(2\sigma_{1}^{2})})\sum
_{N_{M}}(x_{i}-\mu)^{2}}\cdot\frac{1}{(\sqrt{1-\rho^{2}}\sigma
_{2}^{2})^{|\overline{N}_{M}|}}e^{-{J}/{(2(1-\rho^{2})\sigma_{2}^{2})}}.
\end{eqnarray*}

The conditional posterior distributions of all parameters can then
be deduced as follows:
\begin{eqnarray*} &&\sigma_{1}^{2}|
\sigma_{2}^{2},\beta,\mu,\rho,M,X,T\sim \Gamma^{-1}
\biggl(g+\frac{|N_{M}|}{2},h+\frac{1}{2}\sum_{N_{M}}(x_{i}-
\mu)^{2} \biggr),
\\
&&\sigma_{2}^{2}|\sigma_{1}^{2},\beta,
\mu,\rho,M,X,T \sim\Gamma^{-1} \biggl(a+|\overline{N}_{M}|,b+
\frac{J}{2(1-\rho^{2})} \biggr),
\\
&&\beta|\sigma_{1}^{2},\sigma_{2}^{2},
\mu,\rho,M,X,T\\
&&\qquad \sim N \biggl(\frac
{K}{{1}/{s}+({1}/{((1-\rho^{2})\sigma_{2}^{2})})\sum_{\overline
{N}_{M}}(t_{i_{1}}^{2}+t_{i_{2}}^{2}-2\rho t_{i_{1}}t_{i_{2}})},\\
&&\hspace*{17pt}\qquad\quad \frac
{1}{{1}/{s}+({1}/{((1-\rho^{2})\sigma_{2}^{2})})\sum_{\overline
{N}_{M}}(t_{i_{1}}^{2}+t_{i_{2}}^{2}-2\rho t_{i_{1}}t_{i_{2}})} \biggr),
\\
&&\mu|\sigma_{1}^{2},\sigma_{2}^{2},
\beta,\rho,M,X,T \sim  N \biggl(\frac{
{p}/{q}+\sum_{N_{M}}{x_{i}}/{\sigma_{1}^{2}}}{{1}/{q}+
{|N_{M}|}/{\sigma_{1}^{2}}},\frac{1}{{1}/{q}+{|N_{M}|}/{\sigma
_{1}^{2}}} \biggr),
\\
&&\rho|\sigma_{1}^{2},\sigma_{2}^{2},
\beta,\mu,M,X,T\propto\rho ^{u-1}(1-\rho)^{v-1} \biggl(
\frac{1}{\sqrt{1-\rho^{2}}\sigma
_{2}^{2}} \biggr)^{|\overline{N}_{M}|}e^{-{J}/{(2(1-\rho^{2})\sigma
_{2}^{2})}},
\\
&& M|\sigma_{1}^{2},\sigma_{2}^{2},\beta,
\mu,\rho,X,T\\
&&\qquad\propto \biggl(\frac
{1}{\sigma_{1}} \biggr)^{|N_{M}|}e^{-{1}/{(2\sigma_{1}^{2})}\sum
_{N_{M}}(x_{i}-\mu)^{2}}
\biggl(\frac{1}{\sqrt{1-\rho^{2}}\sigma
_{2}^{2}} \biggr)^{|\overline{N}_{M}|}e^{-{J}/{(2(1-\rho^{2})\sigma_{2}^{2})}},
 \end{eqnarray*}
where
\begin{eqnarray*} K&=&\frac{1}{(1-\rho^{2})\sigma_{2}^{2}}\sum
_{\overline
{N}_{M}}\bigl[(t_{i_{1}}-\rho t_{i_{2}})
(x_{i_{1}}-x_{i_{0}})+(t_{i_{2}}-\rho t_{i_{1}})
(x_{i_{2}}-x_{i_{0}})\bigr]+\frac{r}{s},
\\
J&=&\sum_{\overline{N}_{M}}\bigl[(x_{i_{1}}-x_{i_{0}}-
\beta t_{i_{1}})^{2}+(x_{i_{2}}-x_{i_{0}}-\beta
t_{i_{2}})^{2}\\
&&\hspace*{16pt}{}-2\rho (x_{i_{1}}-x_{i_{0}}-\beta
t_{i_{1}}) (x_{i_{2}}-x_{i_{0}}-\beta t_{i_{2}})
\bigr],
\\
&&\hspace*{-31pt}N_{M}\dvtx\mbox{set of cells inside the expression branch with
change point $M$},
\\
&&\hspace*{-31pt}\overline{N}_{M}\dvtx\mbox{set of cells outside the expression
branch with change point $M$}.
\end{eqnarray*}

To fit the change-point-in-tree model in equation \eqref{Model1} to
a tree of cell scores, we use an MCMC algorithm, which iteratively
updates each parameter from its conditional posterior distribution
until converging, as judged by the potential scale reduction factor
(or $\hat{R}$) [\citet{Gelman1992}]. More specifically, an MCMC
chain is said to have converged if $|\hat{R}-1|<0.2$ holds for all
parameters. As shown in \hyperref[suppD]{Supplement D}, this MCMC algorithm
converges fast. The output of the algorithm is a sample from the
converged joint posterior distribution of the change point and all
parameters, from which we can get both the point estimates and the
uncertainty measures of all parameters. More specifically, we
regard the change point value with the highest posterior
probability $M^*$ as the MCMC detected branch, and the conditional
posterior mean values (conditional on the reported $M=M^*$) of
other parameters as fitted parameters.

\subsubsection{Step 3: Use SVR to classify an MCMC detected
branch}\label{step3} The above MCMC algorithm forces the fitting
of a tree of cell scores to the model in equation~\eqref{Model1},
which assumes a single expression branch. Since a target gene may
express in zero or multiple branches in the embryo, the detected
expression branch may be false positive or a nonunique true
positive. To deal with this issue, we feed some features of an
MCMC detected branch to a trained SVR to further decide whether we
shall report the corresponding branch as expressing. SVR is a
version of Support Vector Machine and has been proposed by
\citet{Harris1997}. The used features are provided in
\hyperref[suppC]{Supplement C}. The training of SVR is explained in
Section~\ref{s1}.

If the trained SVR classifies the MCMC detected branch as
expressing, we delete the branch from the tree and run the MCMC
algorithm to fit the change-point-in-tree model again. This
procedure is iterated until an MCMC detected branch is classified
as nonexpression. That is, the trained SVR serves as a criterion
to stop searching more expression branches from the tree of cell
scores.

\subsubsection{Step 4: Find the detailed onset time within a
cell} Steps 2 and 3 report expression at the level of branches in the
tree with cell scores $x_{i}$. Some cells in the SVR reported
branches may not express the target gene. For the cell where the
target gene initiates expressing, the detailed onset time may not
be the birth time of the cell. In step 4, we will further detect
expression onset cells and corresponding times, that is, which data
point $y_{ij}$.

For each 4D confocal movie data file, we make use of the sample mean
$\hat{\mu}$ and sample variance $\hat{\sigma}^{2}$ of the valid data
points except those belonging to expression branches detected at step
3, which provide the most accurate estimation of background noise in
case the tree
contains multiple expression branches. Valid data points which are
greater than the $97.5\%$ quantile of the noise distribution
$N(\hat{\mu},\hat{\sigma}^{2})$ are regarded as extreme
values. We then search along all paths from the change point $M$
to all leaf cells inside the SVR reported branch to identify all
expression segments which satisfy the following: (a) the time series
segment (a
block of neighboring data points) along the path contains at least
10 valid data points; and (b) at least $97.5\%$ percent of the
valid data points in the segment are extreme values. We define a
valid data point as an expression onset if it is the earliest
expression segment time point on a path from the change point $M$
to a leaf cell inside a SVR reported branch, and define a valid
data point as an expression end if it is the last one on a path.

\section{Results and discussion}\label{results}
We use synthesized data, where the background truth is known, to
train SVR and test the performance of our method and compare it with
that of \citet{Murray2012}. We also apply our method on a real data set.

\subsection{Synthesized data}
Three synthesized data sets are generated for simulation studies.

\subsubsection{Synthesized data set 1}\label{s1}
The first data set is synthesized to mimic the real data. To
create a mimic tree of cell scores, we first randomly pick one
well-annotated real tree of cell scores whose expression branches have
been reliably labeled, then use the cell scores of its
nonexpression cells as the background noise distribution to
generate a whole tree of noise cell scores, and finally replace a
random set of branches with real expression branches with the same
branch structures. The above mimicking
procedure is repeated to generate 120 trees of cell scores in
synthesized data set 1. Each of the mimic trees of cell scores
shares the same noise and expression cell score distribution as a
real data file, and we know which cells are really expressing.

The MCMC algorithm in step 2 is run on each of the 120 trees. Once
it converges, the detected branch is deleted and the MCMC
algorithm is run again on the remaining tree, until the MCMC
detected branch no longer contains any really expression cell. In 116
of the 120 trees, the MCMC algorithm
precisely detects all true expression branches first before
finally detecting a nonexpression branch as expressing. It shows
that the MCMC algorithm alone can accurately detect expression
branches from this mimic data set.

\begin{figure}[b]

\includegraphics{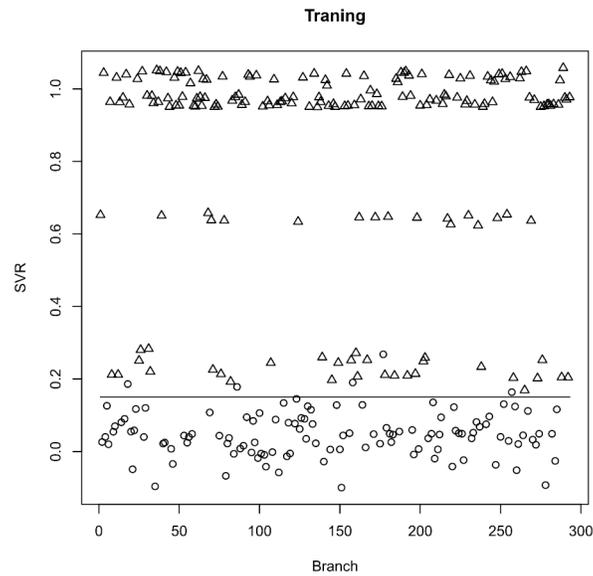}

\caption{Fitted SVR output values of the training branches. Each
point represents a branch, where the horizontal coordinate shows
the branch index and the vertical coordinate shows the SVR output
value. True expression branches are denoted as circles, while
false ones are denoted as triangles. The horizontal line shows the
classification
boundary with SVR output threshold at 0.15.} \label{f3}
\end{figure}

By repeatedly running the above MCMC
algorithm on the 120 mimic trees, the detected branches contain many
true expression
branches (code as output${}={}$1 for SVR) and some false expression
branches (code as output${}={}$0 for SVR). Since we know the true
expression status of these MCMC detected branches, we use selected
features (see \hyperref[suppC]{Supplement C}) of these branches as the training
data set to fit a SVR classifier. Figure~\ref{f3} shows the fitted
output values of all branches in the training data set. The true
expression branches and false expression ones can be fairly
separated by a threshold for the SVR output value. As shown in
\hyperref[suppC]{Supplement~C}, the best threshold is 0.15 because its mean false
classification rate is minimized in this training data set.

To test the accuracy of the trained SVR on independent data, we
synthesize another set of 120 mimic trees and run the MCMC algorithm
using the above same procedure. The trained SVR is used to classify the
MCMC detected branches. Table~\ref{t8} shows the results, where the
false classifications are grouped by the number of true expression
branches in the corresponding trees. Detailed figures are provided in
\hyperref[suppC]{Supplement C} and the mean rate of false classifications is minimized
when the threshold is 0.15, which agrees with the threshold from the
training data. As we can see, the trained SVR performs very well on
this test data.

\begin{table}[t]
\caption{No. of misclassified branches when applying the trained
SVR on MCMC detected branches from testing mimic trees} \label{t8}
\begin{tabular*}{\textwidth}{@{\extracolsep{\fill}}ld{2.0}d{2.0}d{2.0}d{2.0}d{2.0}d{3.0}@{}}
\hline
\multicolumn{1}{@{}l}{\textbf{No. of true expression branches}} &&&&&&\\
\multicolumn{1}{@{}l}{\textbf{contained in corresponding trees}}&\multicolumn{1}{c}{\textbf{None}}&
\multicolumn{1}{c}{\textbf{One}}&
\multicolumn{1}{c}{\textbf{Two}}&\multicolumn{1}{c}{\textbf{Three}}&\multicolumn{1}{c}{\textbf{Four}}&
\multicolumn{1}{c@{}}{\textbf{Overall}}\\
\hline
No. of MCMC detected branches&30&56&81&80&40&287\\
SVR false positive &1&0&0&1&0&2\\
SVR false negative&0&0&0&0&0&0\\
SVR false classifications&1&0&0&1&0&2\\
\hline
\end{tabular*}
\end{table}

\subsubsection{Synthesized data set 2}
The second synthesized data set is generated from the model in
equation \eqref{Model1}, therefore, it fully satisfies all model
assumptions. Using a true data file's lifetime and tree
structure as template, we first randomly select 0 to 10 cells as the
roots of expression branches, then generate true values of
parameters by sampling from their prior distributions, and finally
generate all cell scores according to equation \eqref{Model1}. This
procedure is repeated to generate 110 synthesized trees, with 10
trees for each of the 11 kinds of expression branch numbers.

\begin{figure}[t]

\includegraphics{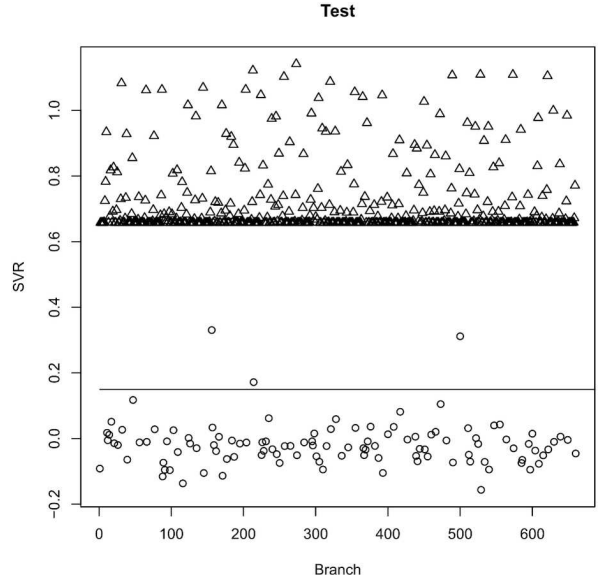}

\caption{Predicted SVR output values for MCMC detected branches
from the second synthesized data set. Triangles indicate true
expression branches, while circles indicate false. The
horizontal line shows the SVR threshold 0.15.}\label{f5}
\end{figure}

For each of the synthesized trees, we run the MCMC algorithm and
use the trained SVR from Section~\ref{s1} to decide when to stop
as described in Sections~\ref{step2} and \ref{step3}. When an MCMC
detected branch is classified as nonexpression by the trained
SVR, the tree is no further fitted by the MCMC algorithm.
Figure~\ref{f5} shows the SVR output values of all SVR reported
branches (triangles above the horizontal threshold line) and the
MCMC detected branches which are classified as nonexpression by
the trained SVR (circles below the horizontal threshold line). It
shows that all true expression branches are correctly reported.
Only three false branches are reported by the trained SVR.

\subsubsection{Synthesized data set 3}
The third data set is used to compare the performance of the DEGEO
algorithm and the method proposed by \citet{Murray2012} (denoted
as APM) in detecting expression onset cells. The data set is
synthesized by mimicking the trees of
original data files. More specifically, we first pick one
well-annotated real tree whose expression onset cells have been
reliably labeled, then use the data points $y_{ij}$ of its
nonexpression branches as the background noise distribution to
generate a whole tree of noise data points, and then insert the
data points $y_{ij}$ of a random set of real expression branches.
The above procedure is repeated to generate 120 trees for the
synthesized data set 3. Each of the mimic trees shares the same
noise and expression distribution as a real data file, and we know
which cells are expression onsets. Table~\ref{tc} shows True Positive
Rate (TPR), False Positive Rate (FPR) and Positive Predictive Value
(PPV) of the two methods at the cell level. The estimated probabilities
under DEGEO have much higher accuracies than those of \citet{Murray2012}.
\begin{table}[b]
\caption{Performance comparison between DEGEO and APM. Standard errors
for APM proportions are approximately 0.034--0.054 for TPR, 0.001--0.002
for FPR, 0.006--0.022 for PPV}\label{tc}
\begin{tabular*}{\textwidth}{@{\extracolsep{\fill}}lcccccc@{}}
\hline
&\multicolumn{2}{c}{\textbf{TPR}}&
\multicolumn{2}{c}{\textbf{FPR}}&
\multicolumn{2}{c@{}}{\textbf{PPV}}\\[-6pt]
&\multicolumn{2}{c}{\hrulefill}&
\multicolumn{2}{c}{\hrulefill}&
\multicolumn{2}{c@{}}{\hrulefill}\\
\multicolumn{1}{@{}l}{\textbf{No. of expression branches}}
&\multicolumn{1}{c}{\textbf{DEGEO}}&\multicolumn{1}{c}{\textbf{APM}}&
\multicolumn{1}{c}{\textbf{DEGEO}}&\multicolumn{1}{c}{\textbf{APM}}&\multicolumn{1}{c}{\textbf{DEGEO}}&
\multicolumn{1}{c@{}}{\textbf{APM}}\\
\hline
0&--&--&0&0.058&--&0\phantom{000.}\\
1&1&0.500&0&0.048&1&0.040\\
2&1&0.528&0&0.047&1&0.086\\
3&1&0.602&0&0.051&1&0.140\\
4&1&0.567&0&0.035&1&0.206\\
\hline
\end{tabular*}
\end{table}

\subsection{Real data}
For the real promoter fusion data from \citet{Murray2012}, the
four-step DEGEO procedure in Section~\ref{methods} is used to find
all expression onset time points.

To evaluate the performance of our method on the real data, we
compile a benchmark real data set by manually annotating
expression branches on 20 real data files. For a comparison with
the SVR stopping criterion on detecting the expression branches,
we also test another intuitive stopping criterion based on the
parameter~$\beta$. More specifically, we stop further MCMC
searching if the $\beta$ of a new MCMC detected branch is less
than one third of the mean values of $\beta$'s of all previously
detected branches. Table~\ref{t3} compares the performances of the
two stopping criteria on the benchmark real data set in terms of
detecting expression branches. It shows that SVR is far better
that the $\beta$-based stopping criterion, because it reports
most of the true expression branches with an acceptable false
negative rate. Detailed results on the benchmark set are provided
in \hyperref[suppC]{Supplement C}.

\begin{table}[t]
\caption{Performances of the 2 stopping criteria on the benchmark
real data set in terms of the number of wrongly or correctly
reported branches}\label{t3}
\begin{tabular*}{\textwidth}{@{\extracolsep{\fill}}lccc@{}}
\hline
\multicolumn{1}{@{}l}{\textbf{Stopping criterion}}&
\multicolumn{1}{c}{\textbf{False positive}}&
\multicolumn{1}{c}{\textbf{True positive}}&
\multicolumn{1}{c@{}}{\textbf{False negative}}\\
\hline
$\beta$-based&31&143&69\\
$\operatorname{SVR}(0.15)$&15&185&27\\
\hline
\end{tabular*}
\end{table}

We run DEGEO for each of the real promoter fusion data files from
\citet{Murray2012}. The detailed results for each data file are
provided in \hyperref[suppF]{Supplement F}, which includes all SVR
reported expression branches, all exact expression onset time
points and all expression segments. Two expression segments are
merged if they are separated by no more than two data points.\looseness=1

DEGEO reports no false positive from the 6 negative control data
files, indicating the good specificity of our
method. For other data files, we try to compare our results with
available results in current literatures. Table~\ref{t7} shows our
results of several genes together with supporting evidences in
current literatures and results obtained by \citet{Murray2008} (denoted
by ROM),
which used an ad hoc threshold to report the expression onset
among all the data sets. It shows that the onset reported in
current literatures and \citet{Murray2008} are also detected by
DEGEO, but DEGEO detects more exact onset times and more onset
locations. Note that \citet{Krause1995} detected disparate onset
times in various expression branches of gene $hlh\mbox{-}1$, which
suggests that iterative runs of step 2 and estimating exact onset
for different paths in step 4 are necessary.\vadjust{\goodbreak}

\begin{table}
\tabcolsep=0pt
\caption{Comparison of expression onset estimation with
current literatures and \citet{Murray2008}. The ``Onset'' columns
list the embryonic stage (in terms of the number of cells at the
onset time) and the cell [named according to the nomenclature in
\citet{Sulston1983}] containing the expression onset. The
``Expression'' columns show which tissues are expressing the
target gene. The $x$ in cell names works as a wildcard character. The
cited papers in the third column provide the source of the
information}\label{t7}
{\fontsize{8}{10}{\selectfont
\begin{tabular*}{\textwidth}{@{\extracolsep{\fill}}lcccccc@{}}
\hline
&\multicolumn{3}{c}{\textbf{Onset (cells stage)}}&\multicolumn{3}{c}{\textbf{Expression (cells)}}\\[-6pt]
&\multicolumn{3}{c}{\hrulefill}&\multicolumn{3}{c@{}}{\hrulefill}\\
\textbf{Gene}&\textbf{Literature}&\textbf{DEGEO}&\textbf{ROM}&\textbf{Literature}&\textbf{DEGEO}&\textbf{ROM}\\
\hline
{end-3}&{28}&{26--28}&{${<}$200}&{Intestine}&{16 intestine}&{20 intestine}\\
&&{Ex}&{Ex}&{[\citet{Maduroa2005}]}&&\\[6pt]
hlh-1&90$+$&133--140,&90&Muscle precursors&16 muscle and&Muscle\\
&C&161--171&&[\citet{Krause1995}]&1 ganglion&\\
&&Cxpx&Cxpx&&&\\[3pt]
&&178--190&180&&8 muscle cells&Muscle\\
&&Dxx&Dxx&&&\\[3pt]
&12--24&51--87&24&Transiently in MS&22 muscle,&Muscle and\\
&MS&Msxxx&MSxx&[\citet{Krause1995}]&3 ganglion,&pharynx\\
&&&&&2 coelomocyte&\\
&&&&&1 mesoderm&\\
&&&&&and 21 pharynx&\\[6pt]
isw-1&Every&87--350&350&Most&Most&Most but \\
&stage&&&[\citet{Andersen2006}]&&not all\\[6pt]
tbx-38&24&45--51&&Descendants of ABa&9 connective tissue,&ABa \\
&ABaxxx&ABaxxxx&&[\citet{Good2004}]&27 hypodermis,& descendents\\
&&&&&97 nerve tissue,&\\
&&&&&48 pharynx&\\
\hline
\end{tabular*}}}
\end{table}

\begin{figure}[t]

\includegraphics{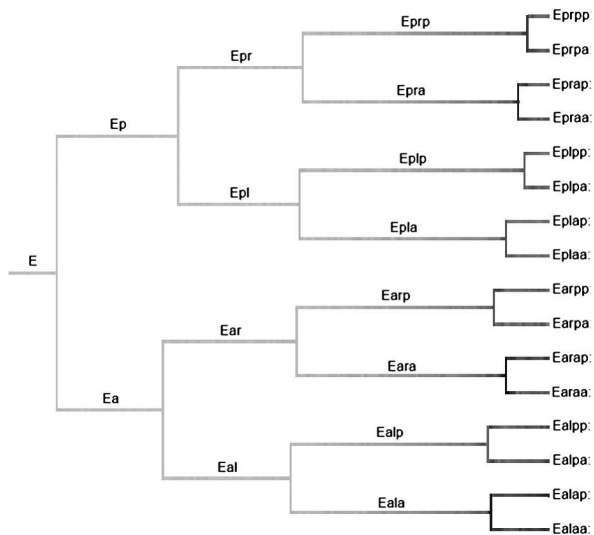}

\caption{The $E$ branch detected from
$\mathit{CD}20070319\_\mathit{pha}4\_I1\mathit{LBBB}.\mathit{csv}$. The expression of the target gene
is found in every path of this expression branch, and the
expression increases faster to high values.}\vspace*{3pt} \label{f10}
\end{figure}

The real data results also show that DEGEO has the capability to
handle the case where the tree contains no expression branch or
more than one expression branch, although the change-point-in-tree
model assumes that the tree contains exactly one expression
branch. DEGEO finds the expression branches one by one, and tends
to first detect the more outstanding expression branch, which
contains more cells and whose expression grows faster to high
values, with a bigger SVR output value. Using the data file
$\mathit{CD}20070319\_\mathit{pha}4\_I1\mathit{LBBB}.\mathit{csv}$ as an example, the $E$ branch shown
in Figure~\ref{f10} is detected with a SVR output value of $0.87$
before the $\mathit{ABarapa}$ branch is detected with a SVR output value of
$0.33$. This tendency is also shown in \hyperref[suppD]{Supplement D}, where
branches with bigger $\beta$ values are detected earlier.\looseness=1

\begin{figure}[b]

\includegraphics{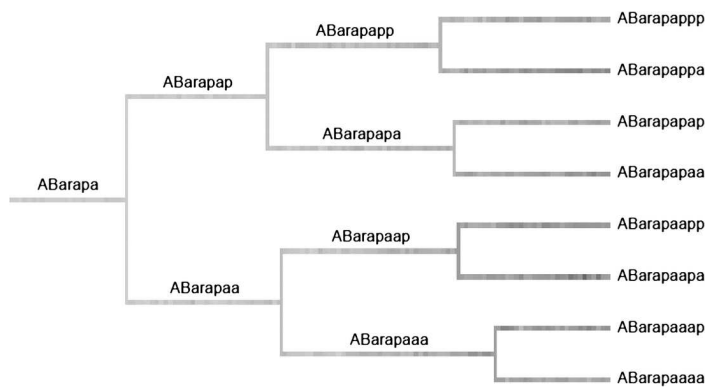}

\caption{The $\mathit{ABarapa}$ branch detected from
$\mathit{CD}20070319\_\mathit{pha}4\_I1\mathit{LBBB}.\mathit{csv}$. The expression of the target gene
is found in every path of this expression branch, but the
expression only increases slowly to relatively low values.}\vspace*{3pt}
\label{f7}
\end{figure}

After the expression branches are detected, DEGEO moves to
estimate the exact expression onset time. For example, DEGEO
reports the cells $\mathit{Exxx}$ and $\mathit{ABarapaxx}$ as expression onset in
Figures \ref{f10} and \ref{f7}, respectively. Here the $x$ in
cell names works as a wildcard character.

\begin{figure}[t]
\centering
\begin{tabular}{@{}cc@{}}

\includegraphics{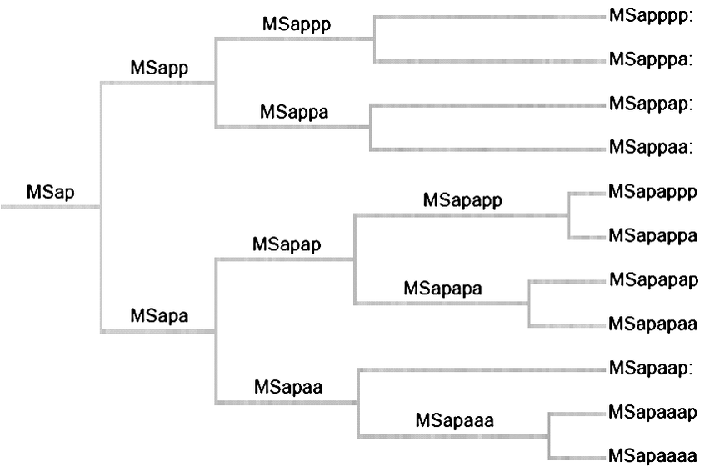}  & \includegraphics{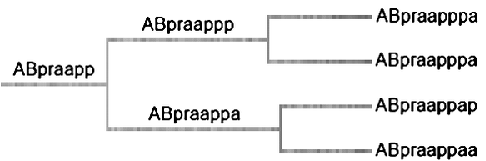}\\
\footnotesize{(A) Msap branch} & \footnotesize{(B) ABpraapp branch}
\end{tabular}
\caption{Examples of false expression phenomenon. \textup{(A)} The $\mathit{MSap}$ branch from
$\mathit{CD}20060627\_\mathit{cnd}1\_4\mbox{-}2.\mathit{csv}$. Its values show a weak uptrend
which probably results from fluorescent absorption. This
branch is classified as a false expression branch by the trained
SVR; \textup{(B)} The $\mathit{ABpraapp}$ branch from
$\mathit{CD}20080128\_\mathit{elt}\mbox{-}1\_3.\mathit{csv}$. Its values show a weak uptrend
which probably results from noise fluctuation. This branch is
classified as a false expression branch by the trained SVR.}
\label{f8}
\end{figure}

DEGEO also seems resistant to the false expression phenomenon
which may result from noise fluctuation or fluorescent absorption.
For example, as shown in Figure~\ref{f8}, the $\mathit{MSap}$ branch from
$\mathit{CD}20060627\_\mathit{cnd}1\_4\mbox{-}2.\mathit{csv}$ and the $\mathit{ABpraapp}$ branch from
$\mathit{CD}20080128\_\mathit{elt}\mbox{-}1\_3.\mathit{csv}$ are correctly classified as
false expression branches by DEGEO.

DEGEO may perform poorly if almost all paths in a tree show
increasing trends, such as in $\mathit{CD}20080504\_C01B7\_1\_6.\mathit{csv}$. This
is because DEGEO assumes all cells outside the selected expression
branch follow a normal distribution, which is invalid if most of
these cells are actually from expression branches. As a result,
the MCMC algorithm will report many expression branches, but may
report relatively weaker expression branches earlier before
stronger expression branches. In this case, we can actually use
step 4 of the DEGEO procedure to directly detect expression onset
on each path without the need to sort out expression branches in
steps 2 and 3.

\section{Conclusion}\label{conclusion}
We provide a principled automatic procedure to detect expression
onset from 4D confocal data of \emph{C. elegans} embryos. Both
simulation studies and real data examples show that our method can
detect both fast and slow expression lineages. On the other hand, it
efficiently excludes false positive ones. Along the paths of
detected expression lineages, we detect exact onset times of the
target gene's expression. Meanwhile, we are able to estimate the
parameters of data files, such as expression rate and distribution
of background noise.

In general, our algorithm can handle most cases well except for
the case where a gene is expressed in almost all cells, because
this case does not fit our model assumption. Extending our model
for multiple change points is a natural choice, but the unknown
number of change points may make the problem computationally very
hard. In this paper, we stick to the assumption of one change
point and test its detection power on the data with multiple
change points. For cases when the gene is not widely expressed,
DEGEO can accurately detect all change points one by one, while
for broadly expressed genes, we come up with a solution by
constructing the background noise distribution from early
expression values instead of all values outside the selected
expression branch.

Except for the embryonic gene onset problem on the cell lineage tree,
our algorithm can also be applied on other change point problems as
long as the data points form a known tree structure. For example, the
information flow on social networks may form such a lineage tree, thus
our algorithm can be used to detect information change points, such as
sentiment formation and propagation [\citet{Liben-Nowell2008}]. The
propagation of contagious disease may also form a lineage tree, and we
can detect the virus mutation on the lineage.

\begin{appendix}\label{app}
\section*{Appendix: Detailed description of the real data}\label{dd}
One fundamental question in biology is how a zygote develops into
an embryo with different tissues. To approach this question,
large-scale 4D confocal movies of \emph{C. elegans} embryos have
been produced by experimental biologists. The first objective is
to detect when and where a gene is expressed in an embryo. Our
real data files are obtained by automated analysis of reporter
gene expression in \emph{C. elegans} with cellular resolution
during embryogenesis [\citet{Murray2012}]. Basically, an embryo is
measured once per minute to report simultaneously the fluorescence
intensity of each cell which is living in the embryo at that time.
Each real data file can be viewed as a binary tree, where each
node is a cell represented by a time series and each branch
indicates a parent--child relationship during cell division. Since
the cell lineage is invariant for all \emph{C. elegans} embryos,
the binary trees from different data files have the same topology.
But a cell's lifetime may vary across different embryos. Overall,
the real data set contains 201 real data files. 5 of them are
negative control files, and each of the remaining files measures
an individual gene's expression during embryogenesis. In total,
111 genes are measured, and 51 genes are measured by replicated
embryos. The $25\%$ quantile, mean, median, $75\%$ quantile and
standard deviation of the distribution of all cell lifetimes are
20, 28.55, 27, 35 and 12.85 minutes, respectively. Some
characteristics of the real data files are summarized in
Table~\ref{ddt}. For more details about the experiment and the
data, please refer to \citet{Bao2006} and \citeauthor{Murray2008}
(\citeyear{Murray2008,Murray2012}).
\end{appendix}

\section*{Acknowledgment}
We thank two anonymous reviewers, the Associate Editor and the
area Editor for their very helpful comments. Supplemental
materials are available online and the R code for the DEGEO
algorithm is available upon request.

\begin{table}[t]
\tabcolsep=0pt
\caption{Distributions of some statistics across the 201 real data
files}\label{ddt}
\begin{tabular*}{\textwidth}{@{\extracolsep{\fill}}ld{5.1}d{5.1}d{5.1}d{5.1}d{5.1}@{}}
\hline
\multicolumn{1}{@{}l}{\textbf{Distribution summary}}&\multicolumn{1}{c}{\textbf{25\% quantile}}&
\multicolumn{1}{c}{\textbf{Mean}}&
\multicolumn{1}{c}{\textbf{Median}}&
\multicolumn{1}{c}{\textbf{75\% quantile}}&\multicolumn{1}{c@{}}{\textbf{Standard
deviation}}\\
\hline
No. of data points&14\mbox{,}369&17\mbox{,}210.4&16\mbox{,}425&20\mbox{,}032&4424.4\\
Observation time (min)&144&173&159&199&39\\
No. of observed cells&708&697.7&713&726&101.5\\
Mean fluorescent intensity&516&7260.2&1834.3&6444.4&13\mbox{,}328.2\\
SD of fluorescent intensity&1714.4&9207&4780.3&12\mbox{,}201.2&10\mbox{,}813.5\\
\hline
\end{tabular*}
\end{table}

\begin{supplement}[id=suppA]
\sname{Supplement A}
\stitle{Model checking}
\slink[doi]{10.1214/15-AOAS820SUPPA} 
\sdatatype{.pdf}
\sfilename{aoas820\_suppa.pdf}
\sdescription{We provide the justification of our 3 model assumptions
in Section~\ref{ass}.}
\end{supplement}

\begin{supplement}[id=suppB]
\sname{Supplement B}
\stitle{Hyperparameters of prior distributions\\}
\slink[doi]{10.1214/15-AOAS820SUPPB} 
\sdatatype{.pdf}
\sfilename{aoas820\_suppb.pdf}
\sdescription{The settings and the sensitivity analysis of
hyperparameters are shown in detail.}
\end{supplement}

\begin{supplement}[id=suppC]
\sname{Supplement C}
\stitle{Classification and stopping criterion based on SVR}
\slink[doi]{10.1214/15-AOAS820SUPPC} 
\sdatatype{.pdf}
\sfilename{aoas820\_suppc.pdf}
\sdescription{We provide plots and tables to demonstrate the good
performance of the SVR method in classifying expression and
nonexpression branches.}
\end{supplement}

\begin{supplement}[id=suppD]
\sname{Supplement D}
\stitle{Convergence diagnosis and parameter estimation}
\slink[doi]{10.1214/15-AOAS820SUPPD} 
\sdatatype{.pdf}
\sfilename{aoas820\_suppd.pdf}
\sdescription{Proofs of successful convergence and good parameter
estimation are provided in additional figures and table.}
\end{supplement}

\begin{supplement}[id=suppE]
\sname{Supplement E}
\stitle{Detection of size-biased sampling\\}
\slink[doi]{10.1214/15-AOAS820SUPPE} 
\sdatatype{.pdf}
\sfilename{aoas820\_suppe.pdf}
\sdescription{We supply some details in detection of the size-biased
sampling problem.}
\end{supplement}

\begin{supplement}[id=suppF]
\sname{Supplement F}
\stitle{Detection results of real data files\\}
\slink[doi]{10.1214/15-AOAS820SUPPF} 
\sdatatype{.zip}
\sfilename{aoas820\_suppf.zip}
\sdescription{All SVR reported expression bran\-ches, all exact
expression onset
time points and all expression segments in real data files are listed
in a folder.}
\end{supplement}


\begin{thebibliography}{23}

\bibitem[\protect\citeauthoryear{Andersen, Lu and Horvitz}{2006}]{Andersen2006}
%
\begin{barticle}[pbm]
\bauthor{\bsnm{Andersen},~\bfnm{Erik~C.}\binits{E.~C.}},
\bauthor{\bsnm{Lu},~\bfnm{Xiaowei}\binits{X.}} \AND
\bauthor{\bsnm{Horvitz},~\bfnm{H.~Robert}\binits{H.~R.}}
(\byear{2006}).
\btitle{\emph{C. elegans} ISWI and NURF301 antagonize an Rb-like pathway in
the determination of multiple cell fates}.
\bjournal{Development}
\bvolume{133}
\bpages{2695--2704}.
\bid{doi={10.1242/dev.02444}, issn={0950-1991}, pii={dev.02444},
pmid={16774993}}
\end{barticle}
%
\bptok{imsref}%
\endbibitem

\bibitem[\protect\citeauthoryear{Bao et~al.}{2006}]{Bao2006}
%
\begin{barticle}[auto:parserefs-M02]
\bauthor{\bsnm{Bao},~\bfnm{Z.}\binits{Z.}},
\bauthor{\bsnm{Murray},~\bfnm{J.~I.}\binits{J.~I.}},
\bauthor{\bsnm{Boyle},~\bfnm{T.}\binits{T.}},
\bauthor{\bsnm{Ooi},~\bfnm{S.~L.}\binits{S.~L.}},
\bauthor{\bsnm{Sandel},~\bfnm{M.~J.}\binits{M.~J.}} \AND
\bauthor{\bsnm{Waterston},~\bfnm{R.~H.}\binits{R.~H.}}
(\byear{2006}).
\btitle{Automated cell lineage tracing in \textit{caenorhabditis elegans}}.
\bjournal{Proc. Natl. Acad. Sci. USA}
\bvolume{103}
\bpages{2707--2712}.
\end{barticle}\vadjust{\goodbreak}
%
\bptok{imsref}%
\endbibitem



\bibitem[\protect\citeauthoryear{Gelman and Rubin}{1992}]{Gelman1992}
%
\begin{barticle}[auto:parserefs-M02]
\bauthor{\bsnm{Gelman},~\bfnm{A.}\binits{A.}} \AND
\bauthor{\bsnm{Rubin},~\bfnm{D.~B.}\binits{D.~B.}}
(\byear{1992}).
\btitle{Inference from iterative simulation using multiple sequences}.
\bjournal{Statist. Sci.}
\bvolume{7}
\bpages{457--472}.
\end{barticle}
%
\bptok{imsref}%
\endbibitem


\bibitem[\protect\citeauthoryear{Good et~al.}{2004}]{Good2004}
%
\begin{barticle}[auto:parserefs-M02]
\bauthor{\bsnm{Good},~\bfnm{K.}\binits{K.}},
\bauthor{\bsnm{Ciosk},~\bfnm{R.}\binits{R.}},
\bauthor{\bsnm{Nance},~\bfnm{J.}\binits{J.}},
\bauthor{\bsnm{Neves},~\bfnm{A.}\binits{A.}},
\bauthor{\bsnm{Hill},~\bfnm{R.~J.}\binits{R.~J.}} \AND
\bauthor{\bsnm{Priess},~\bfnm{J.~R.}\binits{J.~R.}}
(\byear{2004}).
\btitle{The t-box transcription factors tbx-37 and tbx-38 link
glp-1/notch signaling to mesoderm induction in \emph{C. elegans} embryos}.
\bjournal{Development}
\bvolume{131}
\bpages{1967--1968}.
\end{barticle}
%
\bptok{imsref}%
\endbibitem

\bibitem[\protect\citeauthoryear{Guralnik and Srivastava}{1999}]{Guralnik1999}
%
\begin{bincollection}[auto:parserefs-M02]
\bauthor{\bsnm{Guralnik},~\bfnm{V.}\binits{V.}} \AND
\bauthor{\bsnm{Srivastava},~\bfnm{J.}\binits{J.}}
(\byear{1999}).
\btitle{Event detection from time series data}.
In \bbooktitle{KDD'99 Proceedings of the Fifth
ACM SIGKDD International Conference on Knowledge Discovery and Data Mining}
\bvolume{17}
\bpages{33--42}.
\bpublisher{ACM},
\blocation{San Diego, CA}.
\end{bincollection}
%
\bptok{imsref}%
\endbibitem

\bibitem[\protect\citeauthoryear{Harris et~al.}{1997}]{Harris1997}
%
\begin{barticle}[auto:parserefs-M02]
\bauthor{\bsnm{Harris},~\bfnm{D.}\binits{D.}},
\bauthor{\bsnm{Burges},~\bfnm{J.~C.~C.}\binits{J.~C.~C.}},
\bauthor{\bsnm{Kaufman},~\bfnm{L.}\binits{L.}},
\bauthor{\bsnm{Smola},~\bfnm{J.~A.}\binits{J.~A.}} \AND
\bauthor{\bsnm{Vladimir},~\bfnm{N.~V.}\binits{N.~V.}}
(\byear{1997}).
\btitle{Support vector regression machines}.
\bjournal{Adv. Neural Inf. Process. Syst.}
\bvolume{9}
\bpages{155--161}.
\end{barticle}
%
\bptok{imsref}%
\endbibitem


\bibitem[\protect\citeauthoryear{Hu et al.}{2015}]{Hu2015}
%
\begin{bmisc}[author]
{\bauthor{\bsnm{Hu},~\binits{J.}},
\bauthor{\bsnm{Zhao},~\binits{Z.}},
\bauthor{\bsnm{Yalamanchili},~\binits{H.}},
\bauthor{\bsnm{Wang}~\binits{J.}},
\bauthor{\bsnm{Ye},~\binits{K.}} \AND
\bauthor{\bsnm{Fan},~\binits{X.}}}
(\byear{2015}).
\bhowpublished{Supplement to ``Bayesian detection of embryonic gene
expression onset in \emph{C. elegans}.''\newline
DOI:\doiurl{10.1214/15-AOAS820SUPPA},
DOI:\doiurl{10.1214/15-AOAS820SUPPB},\newline
DOI:\doiurl{10.1214/15-AOAS820SUPPC},
DOI:\doiurl{10.1214/15-AOAS820SUPPD},\newline
DOI:\doiurl{10.1214/15-AOAS820SUPPE},
DOI:\doiurl{10.1214/15-AOAS820SUPPF}}.
\bptok{imsref}%
\end{bmisc}
%
\endbibitem
%

\bibitem[\protect\citeauthoryear{Krause}{1995}]{Krause1995}
%
\begin{barticle}[auto:parserefs-M02]
\bauthor{\bsnm{Krause},~\bfnm{M.}\binits{M.}}
(\byear{1995}).
\btitle{Myod and myogenesis in \emph{C. elegans}}.
\bjournal{BioEssays}
\bvolume{17}
\bpages{228}.
\end{barticle}
%
\bptok{imsref}%
\endbibitem

\bibitem[\protect\citeauthoryear{Liben-Nowell and
Kleinberg}{2008}]{Liben-Nowell2008}
%
\begin{barticle}[auto:parserefs-M02]
\bauthor{\bsnm{Liben-Nowell},~\bfnm{D.}\binits{D.}} \AND
\bauthor{\bsnm{Kleinberg},~\bfnm{J.}\binits{J.}}
(\byear{2008}).
\btitle{Tracing information flow on a global scale using Internet
chain-letter data}.
\bjournal{Proc. Natl. Acad. Sci. USA}
\bvolume{105}
\bpages{4633--4638}.
\end{barticle}
%
\bptok{imsref}%
\endbibitem

\bibitem[\protect\citeauthoryear{Liu et~al.}{2009}]{Liu2009}
%
\begin{barticle}[auto:parserefs-M02]
\bauthor{\bsnm{Liu},~\bfnm{X.}\binits{X.}},
\bauthor{\bsnm{Long},~\bfnm{F.}\binits{F.}},
\bauthor{\bsnm{Peng},~\bfnm{H.}\binits{H.}},
\bauthor{\bsnm{Aerni},~\bfnm{S.~J.}\binits{S.~J.}},
\bauthor{\bsnm{Jiang},~\bfnm{M.}\binits{M.}},
\bauthor{\bsnm{Blanco},~\bfnm{A.~S.}\binits{A.~S.}},
\bauthor{\bsnm{Murray},~\bfnm{J.~I.}\binits{J.~I.}},
\bauthor{\bsnm{Preston},~\bfnm{E.}\binits{E.}},
\bauthor{\bsnm{Mericle},~\bfnm{B.}\binits{B.}},
\bauthor{\bsnm{Batzoglou},~\bfnm{S.}\binits{S.}},
\bauthor{\bsnm{Myers},~\bfnm{E.~W.}\binits{E.~W.}} \AND
\bauthor{\bsnm{Kim},~\bfnm{S.~K.}\binits{S.~K.}}
(\byear{2009}).
\btitle{Analysis of cell fate from single-cell gene expression profiles
in \emph{C. elegans}}.
\bjournal{Cell}
\bvolume{139}
\bpages{623--633}.
\end{barticle}
%
\bptok{imsref}%
\endbibitem

\bibitem[\protect\citeauthoryear{Long et~al.}{2009}]{Long2009}
%
\begin{barticle}[pbm]
\bauthor{\bsnm{Long},~\bfnm{Fuhui}\binits{F.}},
\bauthor{\bsnm{Peng},~\bfnm{Hanchuan}\binits{H.}},
\bauthor{\bsnm{Liu},~\bfnm{Xiao}\binits{X.}},
\bauthor{\bsnm{Kim},~\bfnm{Stuart~K.}\binits{S.~K.}} \AND
\bauthor{\bsnm{Myers},~\bfnm{Eugene}\binits{E.}}
(\byear{2009}).
\btitle{A 3D digital atlas of \emph{C. elegans} and its application to
single-cell analyses}.
\bjournal{Nat. Methods}
\bvolume{6}
\bpages{667--672}.
\bid{doi={10.1038/nmeth.1366}, issn={1548-7105}, mid={HHMIMS134260},
pii={nmeth.1366}, pmcid={2882208}, pmid={19684595}}
\end{barticle}
%
\bptok{imsref}%
\endbibitem

\bibitem[\protect\citeauthoryear{Maduroa et~al.}{2005}]{Maduroa2005}
%
\begin{barticle}[auto:parserefs-M02]
\bauthor{\bsnm{Maduroa},~\bfnm{M.~F.}\binits{M.~F.}},
\bauthor{\bsnm{Hillb},~\bfnm{R.~J.}\binits{R.~J.}},
\bauthor{\bsnm{Heidc},~\bfnm{P.~J.}\binits{P.~J.}},
\bauthor{\bsnm{Smitha},~\bfnm{E.~D.~N.}\binits{E.~D.~N.}},
\bauthor{\bsnm{Zhu},~\bfnm{J.}\binits{J.}},
\bauthor{\bsnm{Priess},~\bfnm{J.~R.}\binits{J.~R.}} \AND
\bauthor{\bsnm{Rothman},~\bfnm{J.~H.}\binits{J.~H.}}
(\byear{2005}).
\btitle{Genetic redundancy in endoderm specification within the genus
caenorhabditis}.
\bjournal{Dev. Biol.}
\bvolume{284}
\bpages{522}.
\end{barticle}
%
\bptok{imsref}%
\endbibitem

\bibitem[\protect\citeauthoryear{Murray et~al.}{2008}]{Murray2008}
%
\begin{barticle}[auto:parserefs-M02]
\bauthor{\bsnm{Murray},~\bfnm{J.~I.}\binits{J.~I.}},
\bauthor{\bsnm{Bao},~\bfnm{Z.}\binits{Z.}},
\bauthor{\bsnm{Boyle},~\bfnm{T.~J.}\binits{T.~J.}},
\bauthor{\bsnm{Boeck},~\bfnm{M.~E.}\binits{M.~E.}},
\bauthor{\bsnm{Mericle},~\bfnm{B.~L.}\binits{B.~L.}},
\bauthor{\bsnm{Nicholas},~\bfnm{T.~J.}\binits{T.~J.}},
\bauthor{\bsnm{Zhao},~\bfnm{Z.}\binits{Z.}},
\bauthor{\bsnm{Sandel},~\bfnm{M.~J.}\binits{M.~J.}} \AND
\bauthor{\bsnm{Waterston},~\bfnm{R.~H.}\binits{R.~H.}}
(\byear{2008}).
\btitle{Automated analysis of embryonic gene expression with cellular
resolution in \emph{C. elegans}}.
\bjournal{Nature Methods}
\bvolume{5}
\bpages{703--709}.
\end{barticle}
%
\bptok{imsref}%
\endbibitem

\bibitem[\protect\citeauthoryear{Murray et~al.}{2012}]{Murray2012}
%
\begin{barticle}[auto:parserefs-M02]
\bauthor{\bsnm{Murray},~\bfnm{J.~I.}\binits{J.~I.}},
\bauthor{\bsnm{Boyle},~\bfnm{T.~J.}\binits{T.~J.}},
\bauthor{\bsnm{Preston},~\bfnm{E.}\binits{E.}},
\bauthor{\bsnm{Vafeados},~\bfnm{D.}\binits{D.}},
\bauthor{\bsnm{Mericle},~\bfnm{B.}\binits{B.}},
\bauthor{\bsnm{Weisdepp},~\bfnm{P.}\binits{P.}},
\bauthor{\bsnm{Zhao},~\bfnm{Z.}\binits{Z.}},
\bauthor{\bsnm{Bao},~\bfnm{Z.}\binits{Z.}},
\bauthor{\bsnm{Boeck},~\bfnm{M.}\binits{M.}} \AND
\bauthor{\bsnm{Waterston},~\bfnm{R.~H.}\binits{R.~H.}}
(\byear{2012}).
\btitle{Multidimensional regulation of gene expression in the \emph{C. elegans} embryo}.
\bjournal{Genome Research}
\bvolume{22}
\bpages{1282--1294}.
\end{barticle}
%
\bptok{imsref}%
\endbibitem

\bibitem[\protect\citeauthoryear{Perreault et~al.}{2000}]{Perreault2000}
%
\begin{barticle}[auto:parserefs-M02]
\bauthor{\bsnm{Perreault},~\bfnm{L.}\binits{L.}},
\bauthor{\bsnm{Bernier},~\bfnm{J.}\binits{J.}},
\bauthor{\bsnm{Bobee},~\bfnm{B.}\binits{B.}} \AND
\bauthor{\bsnm{Parent},~\bfnm{E.}\binits{E.}}
(\byear{2000}).
\btitle{Bayesian change-point analysis in hydrometeorological time series}.
\bjournal{Journal of Hydrology}
\bvolume{235}
\bpages{221--241}.
\end{barticle}
%
\bptok{imsref}%
\endbibitem

\bibitem[\protect\citeauthoryear{Picard}{1985}]{Guralnik1985}
%
\begin{barticle}[mr]
\bauthor{\bsnm{Picard},~\bfnm{Dominique}\binits{D.}}
(\byear{1985}).
\btitle{Testing and estimating change-points in time series}.
\bjournal{Adv. in Appl. Probab.}
\bvolume{17}
\bpages{841--867}.
\bid{doi={10.2307/1427090}, issn={0001-8678}, mr={0809433}}
\end{barticle}
%
\bptok{imsref}%
\endbibitem

\bibitem[\protect\citeauthoryear{Spencer et~al.}{2011}]{Spencer2011}
%
\begin{barticle}[auto:parserefs-M02]
\bauthor{\bsnm{Spencer},~\bfnm{W.~C.}\binits{W.~C.}},
\bauthor{\bsnm{Zeller},~\bfnm{G.}\binits{G.}},
\bauthor{\bsnm{Watson},~\bfnm{J.~D.}\binits{J.~D.}},
\bauthor{\bsnm{Henz},~\bfnm{S.~R.}\binits{S.~R.}},
\bauthor{\bsnm{Watkins},~\bfnm{K.~L.}\binits{K.~L.}},
\bauthor{\bsnm{McWhirter},~\bfnm{R.~D.}\binits{R.~D.}},
\bauthor{\bsnm{Petersen},~\bfnm{S.}\binits{S.}},
\bauthor{\bsnm{Sreedharan},~\bfnm{V.~T.}\binits{V.~T.}},
\bauthor{\bsnm{Widmer},~\bfnm{C.}\binits{C.}},
\bauthor{\bsnm{Jo},~\bfnm{J.}\binits{J.}},
\bauthor{\bsnm{Reinke},~\bfnm{V.}\binits{V.}},
\bauthor{\bsnm{Petrella},~\bfnm{L.}\binits{L.}},
\bauthor{\bsnm{Strome},~\bfnm{S.}\binits{S.}},
\bauthor{\bsnm{Stetina},~\bfnm{S.~E.~V.}\binits{S.~E.~V.}},
\bauthor{\bsnm{Katz},~\bfnm{M.}\binits{M.}},
\bauthor{\bsnm{Shaham},~\bfnm{S.}\binits{S.}},
\bauthor{\bsnm{Ratsch},~\bfnm{G.}\binits{G.}} \AND
\bauthor{\bsnm{Miller},~\bfnm{D.~M.}\binits{D.~M.}}
(\byear{2011}).
\btitle{A spatial and temporal map of \emph{C. elegans} gene expression}.
\bjournal{Genome Research}
\bvolume{21}
\bpages{325--341}.
\end{barticle}
%
\bptok{imsref}%
\endbibitem

\bibitem[\protect\citeauthoryear{Sulston et~al.}{1983}]{Sulston1983}
%
\begin{barticle}[pbm]
\bauthor{\bsnm{Sulston},~\bfnm{J.~E.}\binits{J.~E.}},
\bauthor{\bsnm{Schierenberg},~\bfnm{E.}\binits{E.}},
\bauthor{\bsnm{White},~\bfnm{J.~G.}\binits{J.~G.}} \AND
\bauthor{\bsnm{Thomson},~\bfnm{J.~N.}\binits{J.~N.}}
(\byear{1983}).
\btitle{The embryonic cell lineage of the nematode \textit{Caenorhabditis elegans}}.
\bjournal{Dev. Biol.}
\bvolume{100}
\bpages{64--119}.
\bid{issn={0012-1606}, pii={0012-1606(83)90201-4}, pmid={6684600}}
\end{barticle}
%
\bptok{imsref}%
\endbibitem

\bibitem[\protect\citeauthoryear{Yalamanchili et~al.}{2013}]{Yalamanchili2013}
%
\begin{barticle}[auto:parserefs-M02]
\bauthor{\bsnm{Yalamanchili},~\bfnm{H.~K.}\binits{H.~K.}},
\bauthor{\bsnm{Yan},~\bfnm{B.}\binits{B.}},
\bauthor{\bsnm{Li},~\bfnm{M.~J.}\binits{M.~J.}},
\bauthor{\bsnm{Qin},~\bfnm{J.}\binits{J.}},
\bauthor{\bsnm{Zhao},~\bfnm{Z.}\binits{Z.}},
\bauthor{\bsnm{Chin},~\bfnm{F.~Y.}\binits{F.~Y.}} \AND
\bauthor{\bsnm{Wang},~\bfnm{J.}\binits{J.}}
(\byear{2013}).
\btitle{Dynamic delay gene network inference from high temporal data
using gapped local alignment}.
\bjournal{Bioinformatics}
\bvolume{30}
\bpages{377--383}.
\end{barticle}
%
\bptok{imsref}%
\endbibitem
\end{thebibliography}

%





\printaddresses
\end{document}